\documentclass[draftcls,12pt,onecolumn,oneside]{IEEEtran}
\usepackage{mathrsfs}
\usepackage{algorithm}
\usepackage{algorithmic}
\usepackage{url,times,amsmath,color,amssymb,graphicx,epsfig,psfig,cite,geometry,psfrag,subfigure,dsfont,booktabs, multirow}
\begin{document}
\title{Fronthauling for 5G LTE-U Ultra Dense Cloud Small Cell Networks}
\author{Haijun Zhang,~\IEEEmembership{Member,~IEEE}, Yanjie Dong,~\IEEEmembership{Student Member,~IEEE},\\
 Julian Cheng,~\IEEEmembership{Senior Member,~IEEE}, Md. Jahangir Hossain,~\IEEEmembership{Member,~IEEE},\\
  and Victor C. M. Leung,~\IEEEmembership{Fellow,~IEEE}
\thanks{Haijun Zhang is with the Engineering and Technology Research Center for Convergence Networks, University of Science and Technology Beijing, Beijing, China 100083, and also with The State Key Laboratory of Integrated Services Networks, Xidian University (e-mail: dr.haijun.zhang@ieee.org).

Yanjie Dong, Julian Cheng, and Md. Jahangir Hossain are with the School of Engineering, The University of British
Columbia, Kelowna, BC, V1V 1V7 Canada (e-mail: yanjie.dong@alumni.ubc.ca, julian.cheng@ubc.ca, jahangir.hossain@ubc.ca).

Victor C. M. Leung is with the Department of Electrical and Computer Engineering, The University of British Columbia, Vancouver, BC V6T 1Z4 Canada (e-mail: vleung@ece.ubc.ca).

}}
\maketitle

\begin{abstract}
Ultra dense cloud small cell network (UDCSNet), which combines cloud computing and massive deployment of small cells, is a promising technology for the fifth-generation (5G) LTE-U mobile communications because it can accommodate the anticipated explosive growth of mobile users' data traffic. As a result, fronthauling becomes a challenging problem in 5G LTE-U UDCSNet. In this article, we present an overview of the challenges and requirements of the fronthaul technology in 5G \mbox{LTE-U} UDCSNets. We survey the advantages and challenges  for various candidate fronthaul technologies such as optical fiber,  millimeter-wave based unlicensed spectrum, Wi-Fi based unlicensed spectrum, sub 6GHz  based licensed spectrum, and free-space optical based unlicensed spectrum.

\end{abstract}

\section{Introduction}
Driven by the development of mobile Internet and smart phones, data traffic grows exponentially in the current mobile communication systems.
Ultra dense network (UDN) is a promising technique  to meet the requirements of explosive data traffic in the fifth-generation (5G) mobile communications \cite{XiaohuGeUDN2016}.
Moreover, when overlaid on top of the macrocells, low power small cells (such as femtocell and picocell) can improve the coverage and capacity of cellular networks by exploiting spatial reuse of the spectrum \cite{GuandingYuJSAC2015}.
Dense small cells can also offload the wireless data traffic of user equipments (UEs) from macrocells, especially for an indoor environment where more than 80\% of the data traffic occurs.
Since there exists no commonly agreed definition of UDN  in the literature, we define a UDN as a cellular network with traffic volume per area greater than 700 ${\rm{Gbps/}}{{\rm{km}}^2}$ or user density  greater than 0.2 ${\rm{UEs/}}{{\rm{m}}^2}$ in this article.

Since the spectrum is scarce in 5G UDN, a number of researchers have examined to utilize the unlicensed 2.4 GHz and 5 GHz bands used by Wi-Fi for 5G systems, also known as LTE-Unlicensed (LTE-U) systems.
For example, the authors in \cite{GuandingYuJSAC2016} investigated the energy efficiency optimization of licensed-assisted access LTE-U systems.
By leveraging the emerging LTE-U technology, the authors in \cite{GuandingYuTWC2016} proposed that some unlicensed spectrum resources may be allocated to the LTE system in order to compensate more WiFi users.

Besides LTE-U, cloud radio access network (CRAN) is one suitable candidate for 5G systems \cite{mugenCranJSAC2015}.
In a CRAN, baseband processing is centralized in a baseband unit (BBU) pool, and radio frequency  processing is handled in remote radio heads (RRHs).
The BBU pool consists of high performance processors that perform baseband processing functions such as radio resource control, media access control, fast Fourier transform, resource block mapping, modulation, coding, etc.
RRHs provide wireless signal coverage for UEs.
The BBU pool consists of large scale BBUs connected by a high bandwidth, low latency optical fiber network.
Together with system software, BBU pool can constitute a large real time baseband cloud, which is computing cloud in CRAN.
As the baseband signal processing of BBUs is centralized, the CRAN architecture can achieve high processing efficiency and consume less power \cite{HCRANmugenIWC2014}.

It is envisioned that a 5G ultra dense cloud small cell network (UDCSNet) is composed of densely deployed small cells and a CRAN, and such architecture will benefit from both CRAN and small cell networks \cite{HCRANmugenIWC2014}, where radio resource management can be performed efficiently. In addition, BBUs can be removed, added, and upgraded easily in UDCSNet. In a 5G UDCSNet, data transmission and many network functions such as cooperative interference management and mobility handover management require efficient fronthauling between the BBU pool and the RRHs. However, fronthauling can be challenging in 5G UDCSNet for the following reasons: 1) massive deployment of RRHs in UDCSNet makes the  fronthauling complex; 2) traditional optical fiber is expensive for a large scale deployment of RRHs; and 3) huge data traffic requires high fronthaul capacity in UDCSNet.

Many existing works have studied fronthaul\footnote{In this article, fronthaul is defined as the link between RRHs and the BBU pool.} and backhaul\footnote{A backhaul link is traditionally defined as the basestation-to-basestation link or the basestation-to-core-network link. In the CRAN, it is customary to adopt the backhaul link as the link between the BBU pool and the core network.} of 5G systems.
In \cite{5GBackhaul2014}, the authors studied the capacity and energy efficiency of 5G wireless backhaul networks for two typical small cell backhaul solutions. Numerical results suggested a distributed architecture has higher energy efficiency than a centralized architecture.
In \cite{SmallCellBackhaulCommag2013}, the authors investigated the  millimeter-wave (mmWave) and optical fiber backhaul technologies that can satisfy link capacity, coverage, and cost efficiency requirements for future small cells' mobile backhaul.
In \cite{mmWave5GBackhaulCommag2014}, the authors presented a comprehensive tutorial on the use of mmWave frequencies, notably the 60 GHz band and the E band, for small cell access and backhauling as a crucial milestone toward future 5G systems.
The authors identified six key elements in \cite{mmWave5GLiliWeiIWC2014} to enable mmWave communications in future 5G
systems, and  possible solutions were proposed to address the challenges in mmWave communications. The authors in \cite{mmWaveUDNBackhaulCommag2014} pointed out that UDN,  when operating in the millimeter-wave band, is an important component of 5G wireless network.
The authors in \cite{HCRANmugenIWC2015} comprehensively surveyed the recent advances in fronthaul-constrained CRANs, including system architectures and key techniques. Specifically, issues relating to the impact of the capacity-constrained fronthaul on  spectral efficiency, energy efficiency and quality of service for users, including compression and quantization, large-scale coordinated processing, and resource allocation optimization, are discussed together with corresponding potential solutions. However, fronthauling  for 5G UDCSNet has received little attention in the existing literature.

Different from other studies in fronthaul networks, we focus on the fronthaul  networks and consider CRAN-enabled ultra dense  small cell networks for both licensed and unlicensed spectrum.
We first describe a CRAN based ultra dense small cell network with the consideration of phantom cell. Then we evaluate capacity performance of such networks when mmWave is used for both access links (between UEs and RRHs) and fronthaul links.
We also discuss and compare the other potential fronthaul technologies such as unlicensed spectrum\footnote{Hereafter, unlicensed spectrum refers specifically to unlicensed spectrum used by Wi-Fi.} and free space optical (FSO) for 5G UDCSNet.

\section{5G Ultra Dense Cloud Small Cell Network Architecture}
\begin{figure}[h]
        \centering
        \includegraphics*[width=14cm]{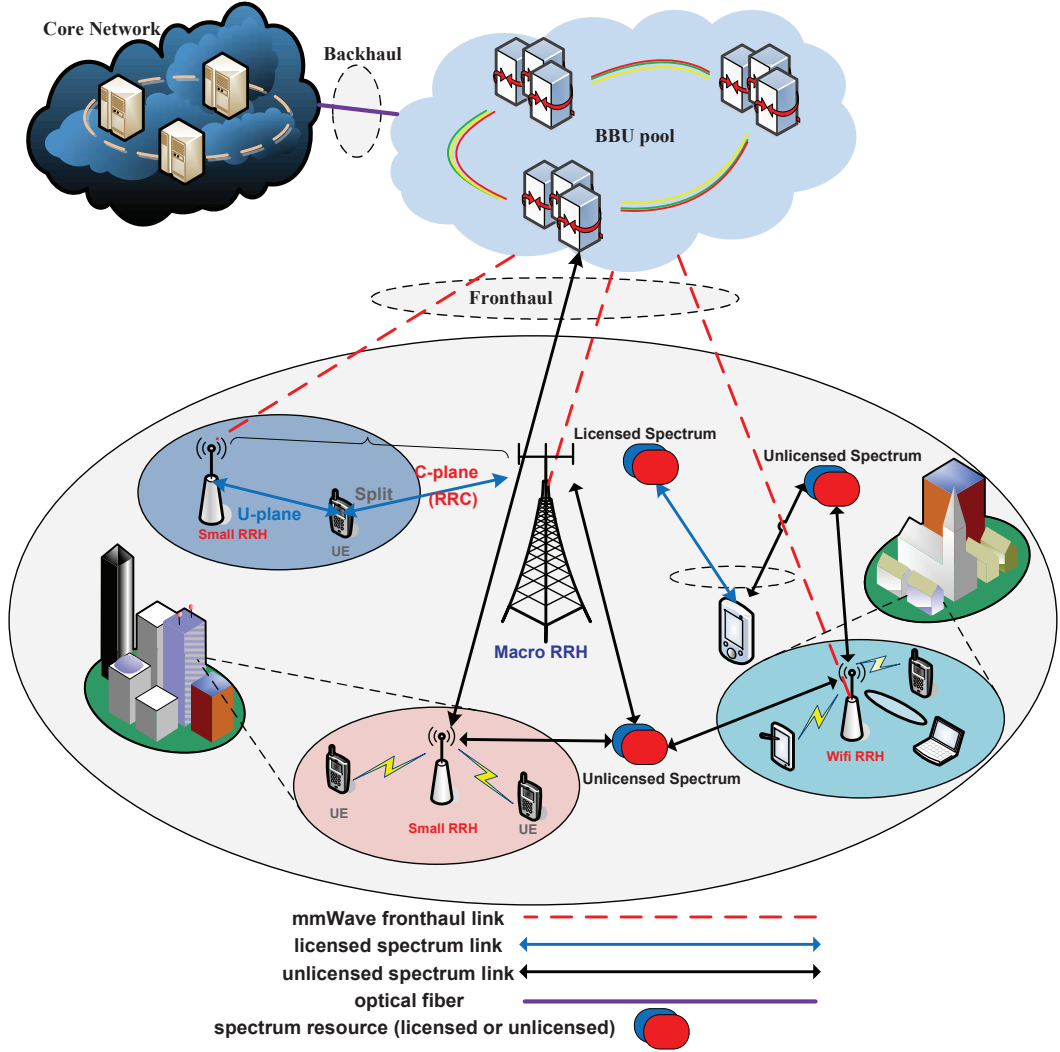}
        \caption{UDCSNet architecture}
        \label{fig:1}
\end{figure}

Figure 1 describes the network architecture of UDCSNet, which is comprised of a macro RRH and many small RRHs, both connected to a BBU pool.

As shown in Fig. 1, the macro RRH and small RRHs act, respectively, like macro basestation (MBS) and small basestations (SBSs) in a heterogeneous network (HetNet). Unlike a traditional HetNet, the radio resource control and management capabilities of MBSs and SBSs are now co-located and processed in the BBU pool.
The small RRHs can be deployed, for example, on each floor of building or office to provide improved capacity and coverage.
The small RRHs in UDCSNet can also be deployed in a hotspot scenario, e.g., in a stadium with ultra dense devices.
MmWave spectrum band can be used for fronthaul links and the access links in UDCSNet, while optical fiber is used for backhaul links between the core network and the BBU pool. The BBU pool in this paper is enhanced with centralized processor and collaborative functions to support the  heterogeneous RRHs (macro RRH, small RRH, Wi-Fi RRH).

In the next section, we will discuss and evaluate the performance of the millimeter-wave based access link and fronthaul link in UDCSNet.

\section{Millimeter-wave based Access Link and Fronthaul Link}

Low power small cells can improve the coverage and capacity of the 5G systems by shortening the distance between transmitter and receiver.
However, the spectrum for cellular network is scarce.
For example, LTE frequency bands cover about only 500 MHz unique spectrum ranging from 699 MHz to 3800 MHz, and many bands are commonly used among LTE, HSPA and GSM.
Besides, not all bands are available in a specific geographic area.
The current spectrum for LTE cannot satisfy the requirement of ultra dense small cell networks. MmWave can be used for radio access and fronthauling in UDCSNet.
The mmWave band ranges from 30 GHz to 300 GHz, which is also called extremely high frequency.
In 30-300 GHz, 60 GHz is more suitable for 5G mobile communications because there is license-free or light-licensed 9 GHz bandwidth available around 60 GHz \cite{mmWave5GBackhaulCommag2014}.
MmWave is an ideal technology to achieve high capacity in UDCSNet  \cite{mmWaveUDNBackhaulCommag2014} due to large available spectrum and high spatial reuse enabled by the narrow directional beam.

\subsection{Millimeter-wave Access Links in an UDCSNet with Phantom Cells}
Small cell and macrocell in UDCSNet can  together be used to form a phantom cell.
As shown in Fig. 1, for the radio protocol architecture of a UE in a phantom cell, its control plane (C-plane) and the user plane (U-plane) are separated.
The name phantom cell is adopted because cell formed by the serving small RRH does not behave like a typical cell (with various control signalings) and it is only designed to carry data traffic.
In a phantom cell, the C-plane is established with the macro RRH in a low frequency band, while the U-plane of the UE is established with its serving small RRH in a high frequency band.
This architecture is contrasting the conventional architecture where both the C-plane and U-plane are provided by the serving macrocell.
Since the C-plane of small cell UEs is managed by the macro RRH, radio resource control (RRC) signalings of small cell UEs are transmitted from the  macro RRH, and handover signaling overhead among the small cells and the macrocell can be significantly reduced, especially for high mobility users.

Since small cells can also be formed by Wi-Fi access points, the phantom cell concept with a split of C-plane and U-plane can also be extended to a Wi-Fi RRH and macrocell scenario shown in Fig. 1. In this case, the C-plane of UEs associated with a Wi-Fi RRH is provided by a macro RRH in a licensed low frequency band, and the U-plane of UEs associated with a Wi-Fi RRH is provided by the serving Wi-Fi RRH in an unlicensed high frequency band.

\begin{figure}[h]
        \centering
        \includegraphics*[width=14cm]{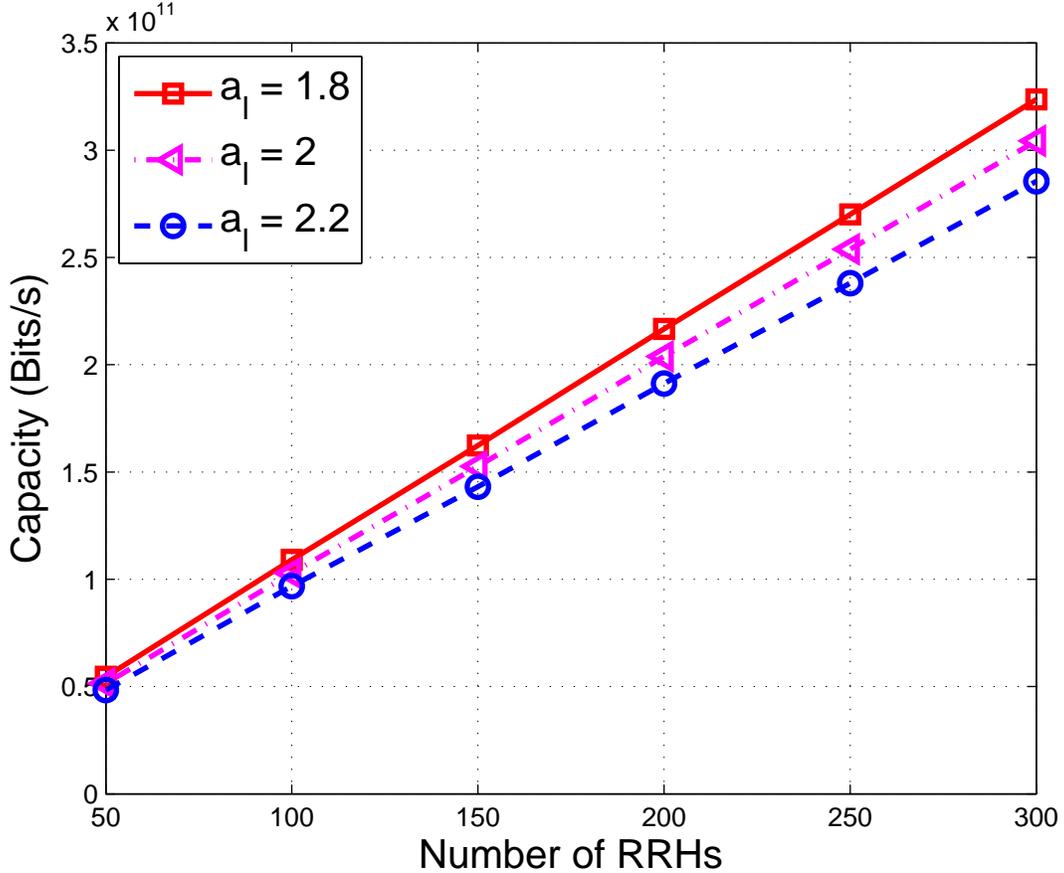}
        \caption{Total capacity of access links in a mmWave-enabled UDCSNet with UE density 0.2 $\mbox{UEs/m}^2$}
        \label{fig:2}
\end{figure}

The use of mmWave  can improve the access link capacity.
Fig. \ref{fig:2} plots the capacity of the total access links in an UDCSNet using 60 GHz mmWave communication.
According to \cite{HurTCOM2013}, we model the path loss as
$PL\left(D\right) = \gamma + 10a_{l}\log_{10}\left(D\right) + \psi$,
where $D$ and $\gamma$ are, respectively, the distance between transmitter and receiver, and the path loss at a close-in reference distance;
$a_l$ is the line-of-sight (LOS) path loss exponent, and the non-line-of-sight (NLOS) components are ignored due to the severe attenuation in the mmWave band;
the term $\psi$ is the fitting derivation.
We model $\psi$  as a Gaussian random variable with zero mean and variance $\sigma^2$.
We assume that both RRHs and UEs are distributed uniformly in a 100 m $\times$ 100 m area with the total number of UEs as 2,000, which yields UE density of 0.2 $\mbox{UEs/m}^2$.
We assume all the UEs share the same channel benefiting from the use of directional beams.
The bandwidth of the channel is 100 MHz, and the noise power is $-174$ dBm/Hz.
The transmit power for each RRH is assumed to be 30 dBm.
As shown in Fig. \ref{fig:2},  the access link capacity of the mmWave enabled UDCSNet can achieve $3.24 \times {10^{11}}$ bps with 300 RRHs and $a_l = 1.8$.
As expected, the capacity of UDCSNet increases as the number of RRH increases.
Because when the number of RRHs increases, UEs can find better RRHs to associate with using the maximum signal-to-noise ratio (SNR) association rule.
Besides, under an ideal transmission scenario, the interference between different users/cells can be neglected due to the use of directional beam.
As shown in Fig. \ref{fig:2}, system capacity with $a_l=1.8$ is higher than that of $a_l=2.2$.
This is because a smaller value of $a_l$ leads to a smaller path loss according to the path loss model.

\subsection{Millimeter-wave for Multihop Fronthaul in UDCSNet}

Since a RRH cannot directly have access to BBU in a NLOS scenario, the RRH can connect with BBU through an aggregation node. The RRH, which is nearest to the BBU pool, can serve as the aggregation node. This aggregation node receives fronthaul traffics from multiple RRHs and forwards them to the BBU pool. The fronthaul link between the aggregation node and the BBU pool can use either optical fiber in NLOS scenario or mmWave in LOS scenario. The RRHs far from the BBU pool can connect to the aggregation node by multihop relaying. The maximum number of hops is considered to be three in order to reduce network delay and routing complexity \cite{mmWaveUDNBackhaulCommag2014}. The RRHs, which act as relay nodes, form a mesh fronthaul network in the UDCSNet.

There are two schemes for mmWave fronthauling: one is out-of-band solution, and the other is in-band solution.
For the in-band solution, the access link and the fronthaul link share the mmWave spectrum. In the out-of-band solution, the fronthaul link uses a different band from the same access link.
The latter approach is more popular in the current microwave cellular networks.
The spectrum of mmWave can be used more efficiently in the in-band solution than that of the out-of-band solution. However, the in-band solution requires effective spectrum reuse plans.
\begin{figure}[h]
        \centering
        \includegraphics*[width=14cm]{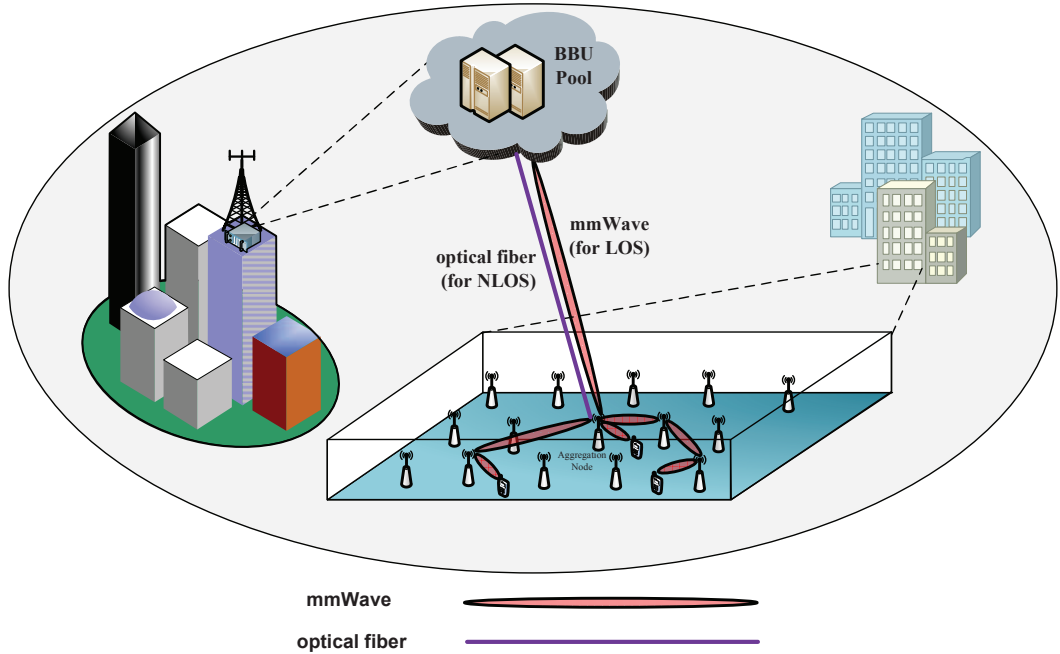}
        \caption{An UDCSNet with multihop mmWave fronthaul}
        \label{fig:multihop}
\end{figure}

There are two  spectrum reuse plans between access link and fronthaul link: 1) frequency division multiplexing, in which access link and fronthaul link use different spectrum bands; 2)  time division multiplexing, in which access link and fronthaul link use the same spectrum in different time slots.
In this case, access link and fronthaul link usually reserve time slots/frames for fronthaul transmission/reception, which is similar to the transmission mode in multihop relays.
Fig. \ref{fig:multihop} illustrates system architecture with multihop mmWave fronthaul in an UDCSNet. As shown from Fig. \ref{fig:multihop}, UEs  can be connected to the aggregation node via a direct link (one hop) or via multi hops. In Fig. 3,  mmWave is used to link RRHs to the aggregation node.

\begin{figure}[h]
        \centering
        \includegraphics*[width=14cm]{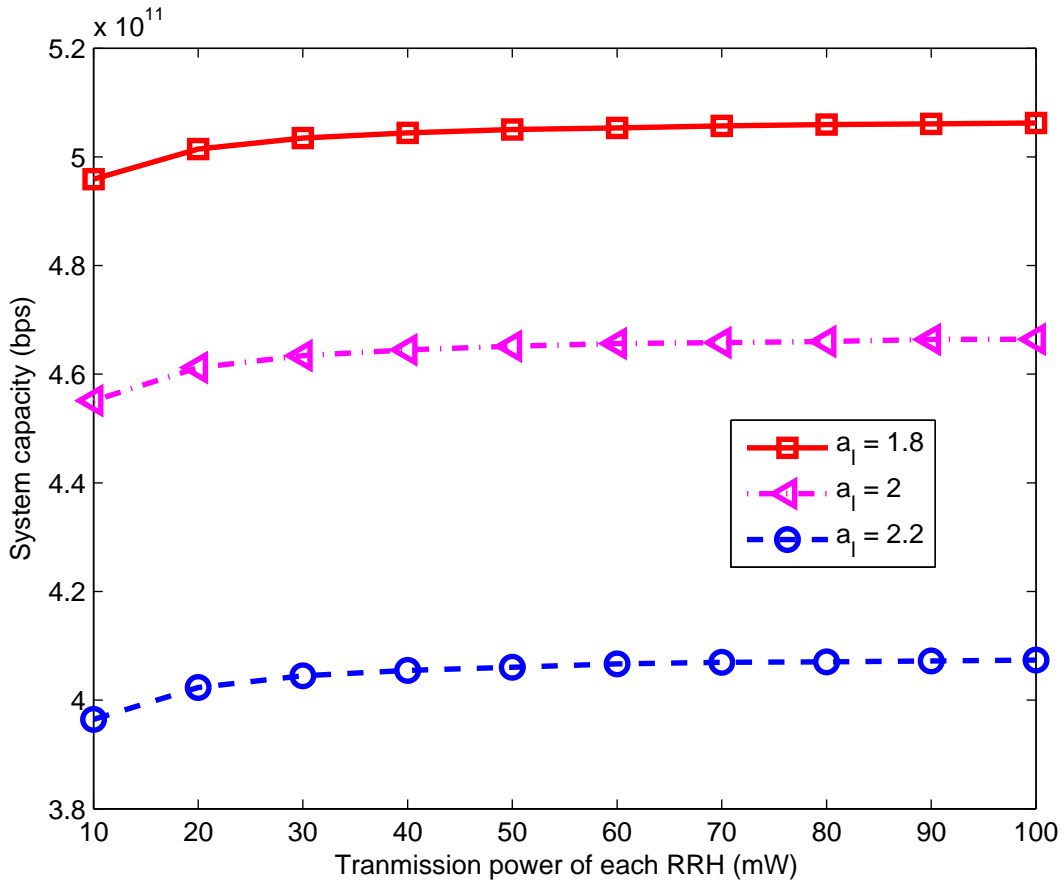}
        \caption{Capacity of the access link and the multihop link of an UDCSNet with UE density 0.2 $\mbox{UEs/m}^2$}
        \label{fig:multihopPerformance}
\end{figure}

Figure \ref{fig:multihopPerformance} shows capacity performance for an UDCSNet assuming all UEs are connected to the aggregation node with two hops.
{
We consider the interference between different RRHs and the rest of the simulation settings follow those of Fig. \ref{fig:2}.
As shown in Fig. \ref{fig:multihopPerformance}, the downlink system end-to-end capacity of the access link and multihop links tends to a stable value.
The reasons are two folds: 1) we consider a practical scenario where interference between  neighboring RRHs and UEs can exist with directional beams with non-negligible sidelobe antenna gain ($-5$ dB); 2) the Shannon capacity formula follows a log-concave feature.
As also shown in Fig. \ref{fig:multihopPerformance}, the system capacity with $a_l=1.8$ is higher than that of $a_l=2.2$.
This is also due to the fact that a smaller value of $a_l$ corresponds to a smaller path loss; and this observation can be inferred from the path loss model. }

\section{Unlicensed Spectrum and Free Space Optical for Fronthauling in 5G UDCSNet}

\subsection{Fronthauling Using Unlicensed Spectrum Access}

Since most of the low frequency bands have been occupied by 2G, 3G, and 4G networks, spectra are being sought in next generation mobile networks.
A number of wireless companies, such as Qualcomm, Huawei, and Ericsson, focus on the LTE-U and propose to use the unlicensed spectrum around 5 GHz frequency band.
However, Wi-Fi has already used unlicensed 2.4 GHz and 5 GHz bands.
The unlicensed spectrum band around 5 GHz is 500 MHz.
The standardization organization (e.g., 3GPP) is examining the co-existence of Wi-Fi and LTE cellular networks.
``Listen Before Talk" makes LTE work effectively with Wi-Fi.
Moreover, ``Listen Before Talk" is  frequency agnostic.

%
%
%

Moreover, the unlicensed spectrum used by Wi-Fi is a potential candidate for fronthaul in UDCSNet. The advantages of using the unlicensed spectrum as a fronthaul are as follows: 1) an operator does not need to purchase separate frequency for fronthaul; 2) the efficiency of the unlicensed spectrum can be improved by reusing in both access links and fronthaul links.

The reuse of unlicensed spectrum between access links and fronthaul links can adopt the following methods: 1) frequency division multiplexing, in which access links and fronthaul links use  different spectrum bands; 2)  time division multiplexing, in which access links and fronthaul links use the spectrum in different time slots; 3) cognitive/opportunistic fronthauling, in which a RRH senses the unlicensed spectrum, and utilizes the unlicensed spectrum as  fronthaul only when the unlicensed spectrum is unused or when the active user signal is  below a predefined interference threshold.

\subsection{Fronthauling Using Free Space Optical}

\begin{figure}[h]
        \centering
        \includegraphics*[width=14cm]{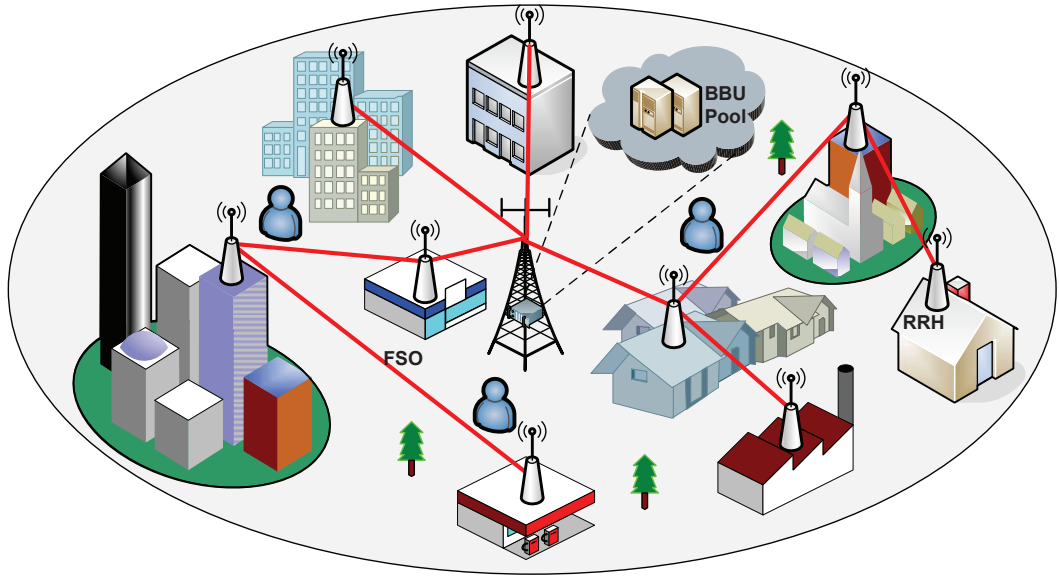}
        \caption{An UDCSNet using FSO for fronthaul link and multihop links.}
        \label{fig:FSO}
\end{figure}

For the fronthaul from RRH to the BBU pool as well as the multihop links, free space optical  communication is a promising candidate complementary to  microwave technology and optical fiber in traditional cellular networks, as described in Fig. \ref{fig:FSO}.

Compared with existing fronthaul solutions, FSO mainly has four advantages: 1) FSO has a wide optical bandwidth, which can result in much higher fronthaul capacity.
Since FSO uses lasers operating in the wavelength range of 800-1700 nm, it can provide several magnitudes of improvement of capacity over the mmWave.
Moreover, the combination of FSO and wavelength division multiplexing can provide even higher capacities in the Gb/s regime \cite{FSOCommag2006}; 2) the laser beam in FSO is directional, so it is  inherently secure and robust  to electromagnetic interference; 3) the frequency band used by FSO is above 300 GHz, and is unregulated; 4) fronthaul based on FSO can be rapidly deployed and installed.

Weather phenomena can affect the performance of FSO, especially for long distance FSO links.
The performance of a FSO link can be degraded sharply in the presence of thick clouds, severe fog, or dust storms \cite{FSOCommag2006}.
For example, in Beijing, where dust storms can be  severe during the spring, fronthauling using FSO may not a good choice.
The poor weather conditions may result in poor transmission performance of FSO.
To achieve both high capacity and link availability, one can use a hybrid FSO/mmWave approach \cite{FSOmmWaveIPC2013}, which can potentially provide carrier grade link availability of 99.999\%.
A hybrid FSO/mmWave system can be robust for either rainy or foggy environment.
While the mmWave performance can be degraded in the presence of rain, it can penetrate fog.
On the other hand, FSO signal using 800-1700 nm lasers cannot be transmitted through thick fog, but the rain will cause little effect on FSO system.
The only weather condition that will affect the performance of a hybrid FSO/mmWave system is the simultaneous heavy rain and thick fog condition, which rarely occurs.

\section{Comparison of Fronthauling Candidates in 5G UDCSNet}
\begin{figure}[h]
        \centering
        \includegraphics*[width=14cm]{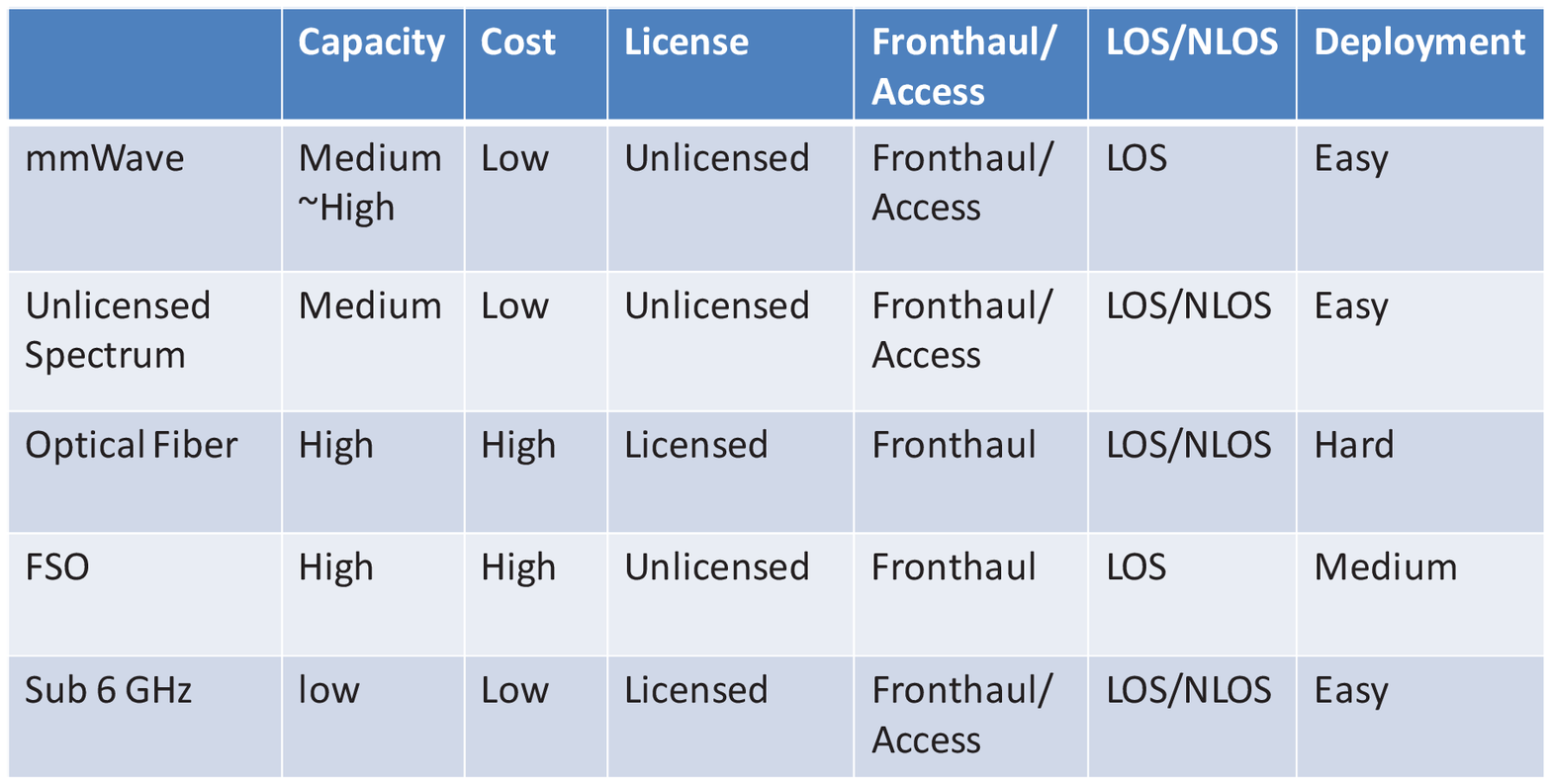}
        \caption{A qualitative comparison of different techniques}
        \label{fig:table}
\end{figure}

Both mmWave and FSO will not operate in a NLOS environment. Instead, optical fiber and sub 6 GHz (licensed) can be used in a NLOS environment.
For LOS fronthaul scenario, FSO, mmWave, unlicensed spectrum will be good choices. Moreover, mmWave and unlicensed spectrum can also be used as the access link.

Figure 6 shows a qualitative comparison of different fronthaul techniques for UDCSNet. We define the cost as capital expenditure (CAPEX) and operational expenditure (OPEX) using difference fronthauling techniques, that is, the cost of equipment, installation, and maintenance etc. The cost of optical fiber depends on the fiber capacity and the distance between BBU and RRH.
The hardware cost of FSO may be high; however, it can be significantly reduced with massive deployment.


\section{Conclusion}
With massive deployment of CRAN-enabled small cells, fronthauling is challenging in LTE-U UDCSNet.
In this article, we provided a summary of recent advances in fronthauling of 5G LTE-U UDCSNet.
To be compatible with the developments of LTE-U and phantom cell in 5G systems, UDCSNet combines advantages of cloud radio access network and ultra dense small cells.
The advantages and challenges of various fronthauling candidates were surveyed in this article.
Specifically, mmWave was examined for both access link and fronthaul link in UDCSNets.
It was found that the interference caused by non-negligible mmWave sidelobe antenna gain can confine the system capacity of a UDCSNet.
Moreover, unlicensed spectrum and free space optical were also investigated for fronthauling in 5G UDCSNet.
Finally, we presented a qualitative comparison  for different fronthauling techniques in UDCSNet.
While the optical based fronthauling techniques are free of interference, the infrastructure cost can be high.
For RF based fronthauling techniques (mmWave, LTE-U, sub 6 GHz), the ongoing challenge remains to be how to cope with the co-channel interference.

\section*{Acknowledgment}
This work was supported by the National Natural Science Foundation of China (61471025)  and the Open Research Fund of The State Key Laboratory of Integrated Services Networks, Xidian University (ISN17-02).

\begin{IEEEbiography}{Haijun Zhang} (M'13) is currently a Full Professor in University of Science and Technology Beijing, Beijing, China. He was a Postdoctoral Research Fellow in in Department of Electrical and Computer Engineering, the University of British Columbia (UBC), Vancouver, Canada. He received his Ph.D. degree in Beijing University of Posts Telecommunications (BUPT). From September 2011 to September 2012, he visited Centre for Telecommunications Research, King's College London, London, UK, as a Visiting Research Associate. He has published more than 70 papers and authored 2 books. He serves as Editor of Journal of Network and Computer Applications, Wireless Networks, Telecommunication Systems, and KSII Transactions on Internet and Information Systems. He also serves as Leading Guest Editor of ACM/Springer Mobile Networks \& Applications (MONET) Special Issue on ``Game Theory for 5G Wireless Networks". He serves as General Chair of GameNets'16, and served as Symposium Chair of the GameNets'14 and Track Chair of ScalCom'15. His current research interests include 5G, Resource Allocation, NOMA, LTE-U, Heterogeneous Small Cell Networks and Ultra-Dense Networks.
\end{IEEEbiography}

\begin{IEEEbiography}{Yanjie Dong}
received the B. Eng. in telecommunication engineering from Xidian University, China, 2011, and the M.A.Sc in Electrical Engineering from UBC, Okanagan, Canada, 2016. He is pursuing his Ph.D. degree in UBC, Vancouver. His research interests focus on the protocol design of energy efficient communications and energy harvesting systems. He was the Webmaster of IEEE Student Branch of UBC, Okanagan campus. He served as the Webmaster of 28th Biennial Symposium on Communications, and Web Committee Chair of 6th International Conference on Game Theory for Networks.
\end{IEEEbiography}

\begin{IEEEbiography}{Julian Cheng} (S'96, M'04, SM'13) received a B.Eng. degree (with first-class honors) in electrical engineering from the University of Victoria, Victoria, BC, Canada, in 1995, a M.Sc.Eng. degree in mathematics and engineering from Queen's University, Kingston, ON, Canada, in 1997, and a Ph.D. degree in electrical engineering from the University of Alberta, Edmonton, AB, Canada, in 2003. He is currently a Full Professor (with tenure) in the School of Engineering, Faculty of Applied Science at The University of British Columbia (Okanagan campus) in Kelowna, BC, Canada. Previously he worked for Bell Northern Research and Northern Telecom (later known as NORTEL Networks). His current research interests include digital communications over fading channels, statistical signal processing for wireless applications, optical wireless communications, and 5G wireless networks.

Dr. Cheng co-chaired the 12th Canadian Workshop on Information Theory (CWIT 2011), the 28th Biennial Symposium on Communications (BSC 2016), and the 6th EAI International Conference on Game Theory for Networks (GameNets 216). He currently serves as an Editor of IEEE COMMUNICATIONS LETTERS, IEEE TRANSACTIONS ON WIRELESS COMMUNICATIONS, and IEEE Access. He served as a Guest Editor for a special issue of the IEEE JOURNAL ON SELECTED AREAS IN COMMUNICATIONS on optical wireless communications. He is also a Registered Professional Engineer in the Province of British Columbia, Canada. Currently, he serves as a Vice President of the Canadian Society of Information Theory.
\end{IEEEbiography}

\begin{IEEEbiography}{Md. Jahangir Hossain} (S'04-M'08) received the B.Sc. degree in electrical and electronics engineering
from Bangladesh University of Engineering and Technology (BUET), Dhaka, Bangladesh; the M.A.Sc. degree from the University of Victoria, Victoria, BC, Canada, and the Ph.D. degree from the University of British Columbia (UBC), Vancouver, BC, Canada.

He served as a Lecturer at BUET. He was a Research Fellow with McGill University, Montreal, QC, Canada; the National Institute of Scientific Research, Quebec, QC, Canada; and the Institute for Telecommunications Research, University of South Australia, Mawson Lakes, Australia. His industrial experiences include a Senior Systems Engineer position with Redline Communications, Markham, ON, Canada, and a Research Intern position with Communication Technology Lab, Intel, Inc., Hillsboro, OR, USA. He is currently working as an Associate Professor in the School of Engineering, UBC Okanagan campus, Kelowna, BC, Canada. His research interests include designing spectrally and power-efficient modulation schemes, quality of service issues and resource allocation in wireless networks. Dr. Hossain regularly serves as a member of the Technical Program Committee of the IEEE International Conference on Communications (ICC). He was an Editor for the IEEE Transactions on Wireless Communications. He received the Natural Sciences and Engineering Research Council of Canada Postdoctoral Fellowship.
\end{IEEEbiography}

\begin{IEEEbiography}{Victor C. M. Leung} (S'75, M'89, SM'97, F'03) is a Professor of Electrical and Computer Engineering and holder of the TELUS Mobility Research Chair at the University of British Columbia (UBC).  His research is in the areas of wireless networks and mobile systems. He has co-authored more than 900 technical papers in archival journals and refereed conference proceedings, several of which had won best-paper awards. Dr. Leung is a Fellow of the Royal Society of Canada, a Fellow of the Canadian Academy of Engineering and a Fellow of the Engineering Institute of Canada. He is serving on the editorial boards of IEEE JSAC-SGCN, IEEE Wireless Communications Letters, IEEE Access and several other journals. He has provided leadership to the technical program committees and organizing committees of numerous international conferences. Dr. Leung was the recipient of the 1977 APEBC Gold Medal, NSERC Postgraduate Scholarships from 1977-1981, a 2012 UBC Killam Research Prize, and an IEEE Vancouver Section Centennial Award.
\end{IEEEbiography}

\end{document}